\documentclass[prb,twocolumn,showpacs,superscriptaddress,preprintnumbers,amsmath,amssymb]{revtex4-1}
\usepackage{graphicx} \usepackage{dcolumn} \usepackage{bm} \usepackage{amsmath} \usepackage{color}
\usepackage{hyperref} \usepackage[utf8]{inputenc} \usepackage[T1]{fontenc}

\begin{document}


\title{Interplay between lattice and spin states degree of freedom in the FeSe superconductor: dynamic spin state instabilities}

\author{Vladimir Gnezdilov}
\affiliation{Institute for Condensed Matter Physics, Technical University of Braunschweig, D-38106 Braunschweig, Germany}
\affiliation{B.I. Verkin Institute for Low Temperature Physics and Engineering, NASU, 61103 Kharkov, Ukraine}

\author{Yurii G. Pashkevich}
\affiliation{Institute for Condensed Matter Physics, Technical University of Braunschweig, D-38106 Braunschweig, Germany}
\affiliation{A.A. Galkin Donetsk Phystech, NASU, 83114 Donetsk, Ukraine}

\author{Peter Lemmens}
\affiliation{Institute for Condensed Matter Physics, Technical University of Braunschweig, D-38106 Braunschweig, Germany}

\author{Dirk Wulferding}
\affiliation{Institute for Condensed Matter Physics, Technical University of Braunschweig, D-38106 Braunschweig, Germany}

\author{Tatiana Shevtsova}
\affiliation{A.A. Galkin Donetsk Phystech, NASU, 83114 Donetsk, Ukraine}

\author{Alexander Gusev}
\affiliation{A.A. Galkin Donetsk Phystech, NASU, 83114 Donetsk, Ukraine}

\author{Dmitry Chareev}
\affiliation{Institute of Experimental Mineralogy, Chernogolovka, Moscow Region 142432, Russia}

\author{Alexander Vasiliev}
\affiliation{Low Temperature Physics and Superconductivity Department, Moscow State University, Moscow 119991, Russia}

\date{\today}

\begin{abstract}

Polarized Raman-scattering spectra of superconducting, single-crystalline FeSe evidence pronounced phonon anomalies with temperature reduction. A large ($\sim 6.5\%$) hardening of the B$_{1g}$(Fe) phonon mode is attributed to the suppression of local fluctuations of the iron spin state with the gradual decrease of the iron paramagnetic moment. The ab-initio lattice dynamic calculations support this conclusion. The enhancement of the low-frequency spectral weight above the structural phase transition temperature $T_s$ and its change below $T_s$ is discussed in relation with the opening of an energy gap between low ($S=0$) and higher spin states which prevents magnetic order in FeSe. The very narrow phonon line widths compared to observations in FeTe suggests the absence of intermediate spin states in the fluctuating spin state manifold in FeSe.

\end{abstract}


\maketitle

\section{Introduction}

The iron selenide FeSe superconductor, being simplest with respect to its crystal structure, is the most unusual compound among the large family of iron based high temperature superconductors (HTSC). In fact, it is the only compound which remains in the nonmagnetic metallic state for temperatures down to 0.02 K.~\cite{khasanov-08} Furthermore, it becomes superconducting in the pure stoichiometry phase,~\cite{mcqueen-09, hu-11} i.e. without doping in contrast to other HTSC members. At first glance it is a conventional superconductor with a modest transition temperature of $T_c \sim 8$ K at ambient pressure.~\cite{hsu-08} However, it shows an enormous growth of $T_c$ up to 37 K at pressures of 9 GPa,~\cite{mizuguchi-08, medvedev-09, margadonna-09} suggesting that the lattice subsystem plays a role in the superconductivity.

From magnetic point of view FeSe and its main building blocks of FeSe$_4$ tetrahedra demonstrate amazing features: A stoichiometric FeSe thin film on a $c$-plane sapphire substrate is a ferromagnetic metal with a Curie temperature above 300 K.~\cite{wu-07} Moreover, in the recently discovered iron selenides intercalated by alkali ions $M_{0.8+x}$Fe$_{1.6+y}$Se$_2$ ($M$ = K, Rb, Cs)~\cite{guo-10, krzton-11, wang-11, fang-11} the FeSe$_4$ tetrahedra in the FeSe layers stay in two different spin states. In the insulating and magnetically ordered part of the phase separated sample the FeSe$_4$ tetrahedra are in the high spin state with a magnetic moment of 3.3 $\mu_B$ per tetrahedron at 10 K.~\cite{ye-11} In the nonmagnetic metallic and superconductive phase of the $M_{0.8+x}$Fe$_{1.6+y}$Se$_2$ system the FeSe$_4$ tetrahedra remain in their low spin state ($S=0$). All of these observations point to an easy switching between the low ($S=0$), intermediate ($S=1$), and high ($S=2$) iron spin states in the layers of FeSe$_4$ tetrahedra and a pronounced instability of the iron spin states. This instability is governed by distortions of the tetrahedra and in particular by the flexible change of the height ($z$-coordinate) of the Se ions above the Fe plane. Thus, it suggests the presence of rather specific longitudinal spin fluctuations in FeSe which are intimately connected to the lattice.

Actually, nonuniform, $q\neq0$, antiferromagnetic spin fluctuations in FeSe, seen in NMR experiments,~\cite{imai-09} are reported to be strong and increasing near the superconducting transition temperature. Furthermore, applied pressure enhances these spin fluctuations together with an increasing $T_c$.~\cite{imai-09} However, the uniform $q=0$ paramagnetic spin susceptibility, which is derived from the NMR Knight shift, shows an almost linear decrease under temperature lowering.~\cite{imai-09} This observation can be interpreted as an unexpected suppression of the uniform spin fluctuations in FeSe with decreasing temperature.

The temperature evolution of spin fluctuations should be seen in the lattice dynamics through a high sensitivity of the phonon energies on the varying iron spin states.~\cite{hahn-09, reznik-09, boeri-10} Already earlier first principle studies for ``1111'' and ``122'' systems have revealed a strong dependence of the $z$-atom positions from the Fe spin state.~\cite{yildirim-09} It was shown that under external pressure, the high Fe-spin state structure should transform to a structure with low Fe-spin state and significantly reduced $c$-axis. Recent experimental and theoretical studies of the zone center phonons in the isostructural FeTe -- a magnetic and nonsuperconductive relative of FeSe -- have shown that the real magnetic moment of the Fe ions should be included in spin polarised lattice dynamic calculations to achieve an agreement between theory and experiment.~\cite{gnezdilov-11, pashkevich-12} Furthermore, the high frequency part of phonon spectra which in the iron chalcogenides HTSC mostly consists of iron vibrations shows a giant high frequency shift under a decrease of the iron magnetic moment. The underlying physics of this phenomenon is related to strong orbital reordering under a change of the spin state.

Here we report results from Raman light scattering and ab-initio lattice dynamic calculations of recently grown high quality FeSe single crystals. We demonstrate that a dynamical spin state instability is realized in FeSe at high temperatures. Strong dynamical fluctuations of the iron spin state are seen as a renormalization of the B$_{1g}$(Fe) phonon energy at high temperatures with a dominance of the high spin state. Under temperature lowering these fluctuations decrease due to the appearance of an energy gap between different spin states with mainly low spin state presence. This is observed as an additional phonon frequency hardening which highly exceeds the usual temperature dependent hardening given by the Gr\"{u}neisen law. The presence of mainly low and high spin fluctuating states in the paramagnetic ground state in FeSe is seen as a one order narrowing of the phonon linewidth compared to the width of those modes in FeTe where a highly orbital frustrated intermediate spin state determines its properties.~\cite{gnezdilov-11} The suppression of spin state fluctuations is continuous and smooth with lowering temperatures. Below 150 K we observe a highly polarized low frequency background which becomes a gap-like structure. This feature is attributed to a gap opening between a manifold of intermediate spin states and the paramagnetic ground state in FeSe. Our findings shed light on the mechanism that prevents magnetic order and the appearance of a superconducting state in FeSe.

\section{Experimental details}

We have studied FeSe single crystals grown in evacuated quartz ampoules using a KCl/AlCl$_3$ flux.~\cite{lin-11} The very well characterized samples were from the same series as in previous investigations.~\cite{lin-11, luo-12} The present samples composition is FeSe$_{0.96}$. Raman spectra were measured on freshly cleaved surfaces in a quasi-back scattering configuration using a $\lambda = 532.1$ nm solid-state laser. The laser power was set to 5 mW with a spot diameter of approximately 100 $\mu$m to avoid heating effects. All measurements were carried out in a closed-cycle cryostat (Oxford/Cryomech Optistat) in the temperature range from 7 to 295 K. The spectra were collected via a triple spectrometer (Dilor-XY-500) by a liquid nitrogen cooled CCD (Horiba Jobin Yvon, Spectrum One CCD-3000V). The polarization configuration of the Raman spectra is denoted by ($e_ie_s$). We used $a$- and $b$-notation for the tetragonal crystallographic axes and $x$- and $y$-notations for the orthorhombic axes which are rotated by $45^\circ$ from the $a$- and $b$-axes.

\section{Results}

At room temperature, FeSe has the tetragonal PbO structure (space group $P4/nmm$).~\cite{schuster-79} In this phase Fe and Se ions occupy 2a and 2c Wyckoff positions, respectively. Symmetry analysis shows that there are four Raman-active modes [$\Gamma_{Raman}$ = A$_{1g}$(Se) + B$_{1g}$(Fe) + 2E$_g$(Se, Fe)] and two infrared-active modes [$\Gamma_{IR}$ = A$_{2u}$(Se, Fe) + E$_u$(Se, Fe)]. The Raman tensors are given by:

\begin{center}
\mbox{A$_{1g}$=$\begin{pmatrix} a & 0 & 0\\ 0 & a & 0\\ 0 & 0 & b\\
\end{pmatrix}$

, B$_{1g}$=$\begin{pmatrix} c & 0 & 0\\ 0 & -c & 0\\ 0 & 0 & 0\\
\end{pmatrix}$

, and E$_{g}$=$\begin{pmatrix} 0 & 0 & -e\\ 0 & 0 & e\\ -e & e & 0\\ \end{pmatrix}$.}
\end{center}

\begin{figure}
\centering
\includegraphics[width=8cm]{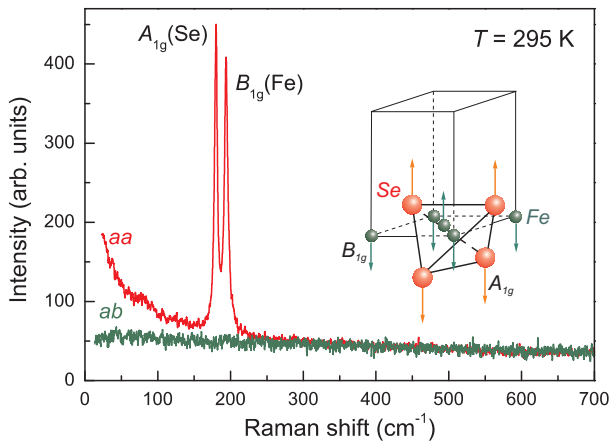}
\caption{\label{295K-overview}(Color online) Polarized Raman spectra of FeSe at room temperature. The inset shows the atomic displacement patterns of the Raman-active phonon modes.}
\end{figure}

Indeed, two sharp peaks are observed in the tetragonal phase at room temperature at 179.8 cm$^{-1}$ [A$_{1g}$ (Se)] and 193.9 cm$^{-1}$ [B$_{1g}$(Fe)], see the ($aa$) spectrum in Fig.~\ref{295K-overview}. On the one hand, the narrow linewidth of the peaks indicates a high quality of the samples. On the other hand, a monotonic phonon mode sharpening was observed in Fe$_{1+y}$Te$_{1-x}$Se$_x$ with increasing Se content.~\cite{um-12} This fact is quite surprising since one may expect a concentration dependent broadening of the phonon lineshape due to the Se substitution-induced disorder. It was argued that spin-phonon coupling may be responsible for this behaviour.

At $T_s = 90$ K, FeSe undergoes a second-order ferroelastic phase transition from the tetragonal to an orthorhombic phase, and below 90 K, its space symmetry group is $Cmma$, with a $\sqrt{2} \times \sqrt{2}$ supercell enlargement in the basal plane.~\cite{margadonna-et-al} In the low-temperature orthorhombic phase Fe and Se ions occupy $4a$ and $4g$ Wyckoff positions, respectively. Symmetry analysis gives six Raman-active modes [$\Gamma_{Raman}$ = A$_g$(Se) + B$_{1g}$(Fe) + 2B$_{2g}$(Se,Fe) + 2B$_{3g}$(Se,Fe)] and six IR-active modes [$\Gamma_{IR}$ = 2B$_{1u}$(Se,Fe) + 2B$_{2u}$(Se,Fe) + 2B$_{3u}$(Se,Fe)]. The corresponding Raman tensors are given by:

\begin{center} \begin{widetext}
\mbox{A$_{g}$=$\begin{pmatrix} a & 0 & 0\\ 0 & b & 0\\ 0 & 0 & c\\
\end{pmatrix}$

, B$_{1g}$=$\begin{pmatrix} 0 & d & 0\\ d & 0 & 0\\ 0 & 0 & 0\\
\end{pmatrix}$

, B$_{2g}$=$\begin{pmatrix} 0 & 0 & e\\ 0 & 0 & 0\\ e & 0 & 0\\
\end{pmatrix}$

, and B$_{3g}$=$\begin{pmatrix} 0 & 0 & 0\\ 0 & 0 & f\\ 0 & f & 0\\ \end{pmatrix}$.} \end{widetext}
\end{center}

\begin{figure}
\centering
\includegraphics[width=8cm]{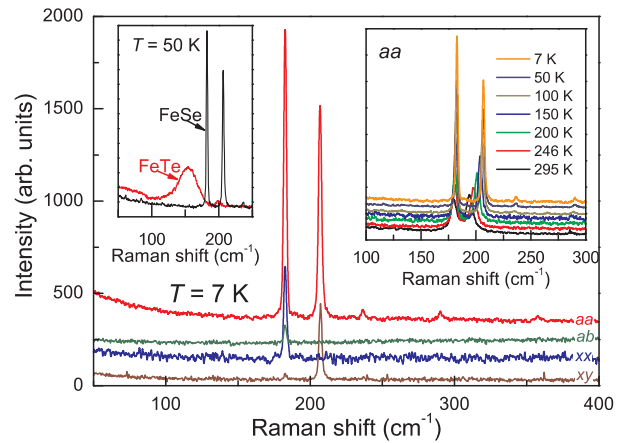}
\caption{\label{7K-overview}(Color online) Polarized Raman spectra of FeSe at $T = 7$ K. The left inset shows a comparison between FeSe and FeTe at $T=50$ K. The right inset shows the temperature evolution of the ($aa$) polarized spectra.}
\end{figure}

Figure~\ref{7K-overview} presents polarized Raman spectra of FeSe at $T = 7$ K. The A$_{1g}$ mode in the tetragonal structure changes into an A$_g$ mode in the orthorhombic structure by the compatibility relation of the group theory and is observable at 182.6 cm$^{-1}$ in ($aa$) and ($xx$) scattering configurations. The B$_{1g}$ mode at 206.9 cm$^{-1}$ is observed in ($aa$) and ($xy$) scattering geometry. Besides, weak lines with frequencies around 236, 290 and 357 cm$^{-1}$ appear in ($xx$) polarized spectra at low temperatures, which can be assigned to B$_{2g}$/B$_{3g}$ modes. The frequencies of these modes have no counterparts in Raman active phonon modes of FeSe, therefore we assign them to a small admixture of iron/selenium oxides.~\cite{hu-11} Here we want to highlight that the positions of the phonon lines in our investigations do not coincide with those of a Raman study in Ref.~\cite{kumar-10}. The reason may be related to sample quality of the previous investigation. A much better agreement is found with the data of a phonon density of states study by $^{57}$Fe nuclear inelastic scattering.~\cite{ksenofontov-10} Here the B$_{1g}$(Fe) phonon mode was identified at the frequency of 196.8 cm$^{-1}$.

\begin{figure*}
\centering
\includegraphics[width=12cm]{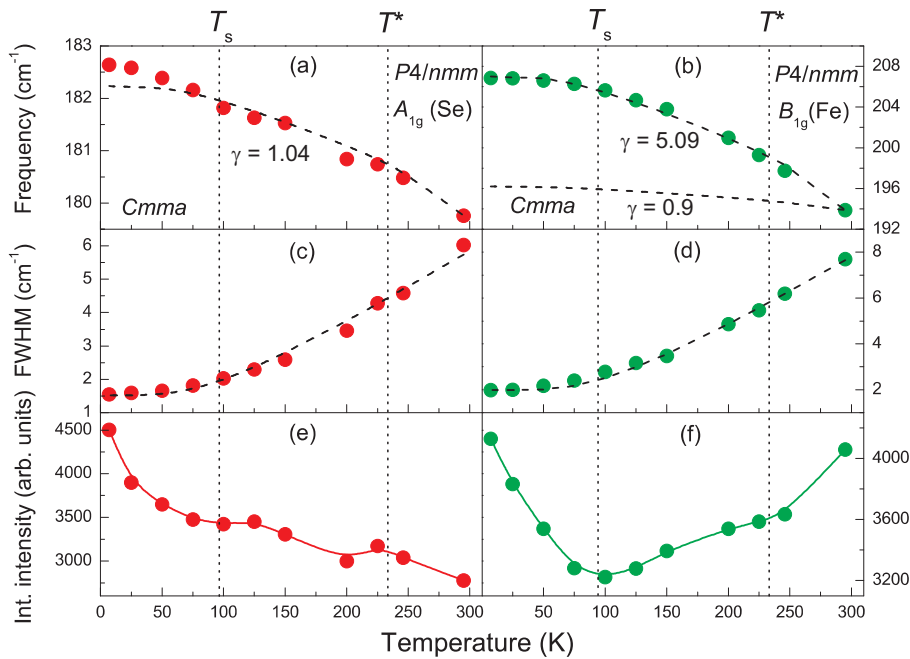}
\caption{\label{fitparameters}(Color online) Temperature dependence of the A$_{1g}$(Se) and B$_{1g}$(Fe) phonon line parameters in FeSe. (a, b) -- phonon frequencies. Black dashed lines correspond to Gr\"{u}neisen dependencies with $\gamma_{A1g(Se)} = 1.04$, $\gamma_{B1g(Fe)} = 5.09$, and $\gamma_{B1g(Fe)} = 0.9$. (c, d) -- linewidths (FWHM). Black dashed lines correspond to fits with Eq. (1). (e, f) -- integrated intensities.}
\end{figure*}

The temperature evolution of the ($aa$) phonon spectra is presented in the right inset of Fig.~\ref{7K-overview}. To get more insight into the phonon dynamics, we analyzed the temperature dependence of the phonon lines parameters. Lines were fitted with Lorentzians in the whole investigated temperature range and results are presented in Fig.~\ref{fitparameters} where the temperature dependent parameters of the A$_{1g}$ and B$_{1g}$ phonon modes are plotted.

The $T$ dependencies of the frequencies of both modes are smooth, see Fig.~\ref{fitparameters} (a) and (b), their positions shift to a higher energy with the temperature decreasing. The A$_{1g}$(Se) mode hardens by 3 cm$^{-1}$ while the B$_{1g}$(Fe) mode undergoes a large hardening by 14 cm$^{-1}$ (6.5\%) in the whole temperature interval. An analysis shows that such a behaviour can not be explained only by the crystal lattice contraction upon cooling. Using the data for the lattice parameters from Ref.~\cite{mizuguchi-10}, we can estimate that in the frame of the Gr\"{u}neisen law, the parameter $\gamma_i$ must be equal to $\sim 1.04$ for the A$_{1g}$(Se) mode and $\sim 5.09$ for the B$_{1g}$(Fe) mode. Both of these values deviate from the conventional value $\gamma \sim 2$ as well as from $\gamma_{A1g(Te)} = 2.85$ and $\gamma_{B1g(Fe)} = 2.15$ as observed by us in FeTe.~\cite{gnezdilov-11}
A hardening of the phonon density of state in the region of the zone centred B$_{1g}$(Fe) phonon energy was observed in Ref.~\cite{ksenofontov-10} and can be estimated with a Gr\"uneisen parameter of $\gamma_{B1g(Fe)} \sim 4$.

The first value is quite reasonable. However, the latter deviates from the conventional values. The hardening of all phonon modes under pressure was observed in Ref.~\cite{ksenofontov-10} and the estimated Gr\"{u}neisen parameter is $\gamma_{B1g(Fe)} = 0.9$. Given that only a part of the observed frequency shift can be attributed to 1\% of the unit cell volume thermal contraction~\cite{mcqueen-09, phelan-09} and to an increasing anharmonicity of the vibrational modes with increasing temperature,~\cite{litvinchuk-08} we believe that the anomalous B$_{1g}$(Fe) phonon hardening is connected with the spin-state of Fe ions changing as well.

The solid lines in the Fig.~\ref{fitparameters} (c) and (d) are a fit of the temperature dependence of the phonon linewidths using
\begin{equation}
\Gamma(T) = \Gamma_0 \left( 1 + \frac{d_i}{\exp(\hbar \omega_0 / k_B T)-1} \right)
\end{equation}
with $\Gamma_0 = 1.52/1.98$ cm$^{-1}$, the eigenfrequency $\omega_0 = 182.6/206.9$ cm$^{-1}$, and $d_i$ being a mode dependent fit parameter for A$_{1g}$/B$_{1g}$ phonon modes. The data can be described very well down to the lowest temperature by this expression, which includes cubic anharmonicity effects.~\cite{klemens-66, balkanski-83} Noteworthy, however, is the superimposed change of curvature leading to an additional drop of linewidth for temperatures around $T_s$ for the B$_{1g}$ phonon mode.

Figures~\ref{fitparameters} (e, f) present the integrated intensity of two phonon lines, in analogy to the frequency and linewidth of the phonon modes. It is noteworthy that the intensity evolution is different for each mode. However, both curves have features which are consistent with the temperature of the structural phase transition, $T_s = 90$ K, and the temperature $T^* = 230$ K, at which a new feature, associated with changes in the structural sector, has been found in studies of Ref.~\cite{luo-12}. Evidently also, the A$_{1g}$(Se) mode reveals a regular deviation from a purely Gr\"{u}neisen behaviour for $T < 90$ K (Fig.~\ref{fitparameters} (a)).

\begin{figure}
\centering
\includegraphics[width=8cm]{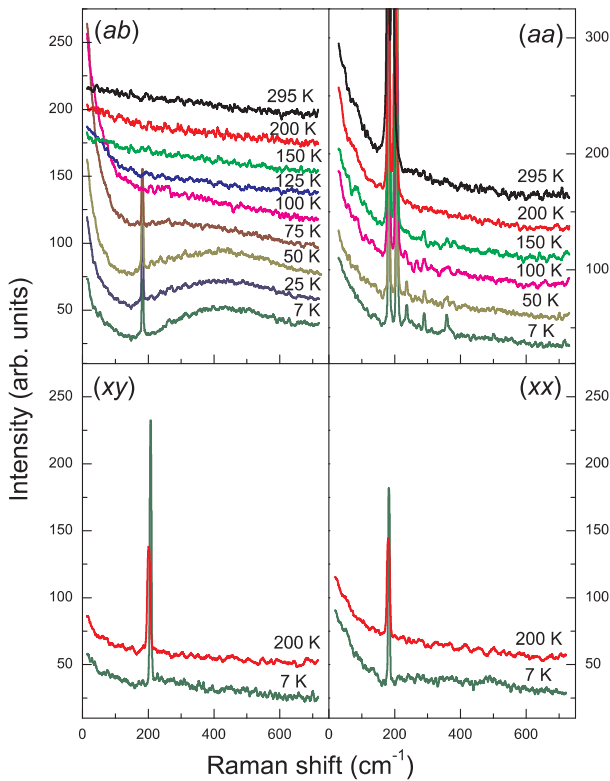}
\caption{\label{polarization}(Color online) Temperature dependent Raman spectra of FeSe. The low energy quasielastic spectral weight does not show a pronounced temperature evolution in the ($aa$), ($xx$) and ($xy$) channels of scattering. In the ($ab$) channel (B$_{2g}$ channel in the tetragonal setting) quasielastic scattering is not present at high temperature but develops below 150 K.}
\end{figure}

As mentioned above, enhanced spin fluctuations at the nesting vector may cause a spin-density wave. A previous Raman scattering study of BaFe$_2$As$_2$~\cite{sugai-12} revealed changes of the electronic properties at the SDW transition. The nodal and anti-nodal gap excitations with different symmetry were observed in the spectra and theoretically analyzed. In order to clarify the trend in going from Fe-As based materials to FeSe, we performed a detailed study of polarization and temperature dependence of our spectra. The temperature dependence of the Raman spectra in ($aa$), ($ab$), ($xx$), and ($xy$) polarization configurations is shown in Fig.~\ref{polarization}.

\begin{figure}
\centering
\includegraphics[width=8cm]{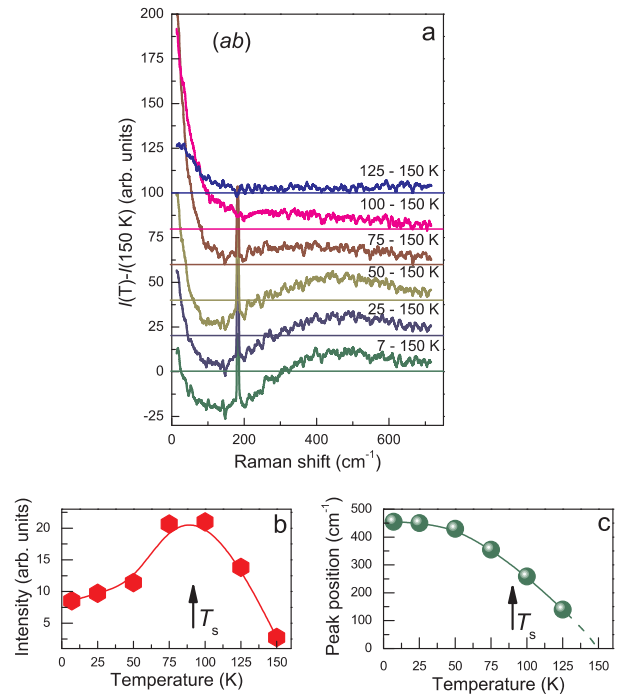}
\caption{\label{differential}(Color online) (a) Differential Raman spectra from 150 K in ($ab$) polarization. The zero levels are shown by the horizontal lines with the same color as the spectra. (b) Integrated intensity of the quasi-elastic tail. (c) Energy of the hump as function of temperature.}
\end{figure}

At all temperatures the spectra have a low-energy quasi-elastic spectral weight except in ($ab$) scattering geometry, where the quasi-elastic signal appears only at temperatures below $\sim 150$ K. However, the temperature dependence of the low-frequency part is very different for ($aa$), ($ab$), ($xx$), and ($xy$) polarizations. In ($aa$) scattering geometry, the quasi-elastic signal only slightly decreases with temperature reduction, in ($xy$) it is nearly constant, and in ($xx$) it has a slightly different shape at low temperatures. Drastic changes are observed in the ($ab$) geometry. The low-energy scattering intensity abruptly increases at temperatures below 150 K and then a gap-like structure forms together with a very broad hump at 350 cm$^{-1}$ below $T_s$. Simultaneously, the low-energy quasi-elastic intensity starts to evolve. Beside, a line with the frequency 182 cm$^{-1}$ appears in the ($ab$) spectra below $T_s$ due to orthorhombic distortions and the appearance of twins as well as small twins rotations in $ab$-plane. Figure~\ref{differential} (a) shows difference spectra between various temperatures and 150 K. The zero levels are indicated by the horizontal lines with the same color as the spectra. Fig.~\ref{differential} (b) demonstrates a temperature dependence of the integrated intensity of the quasi-elastic tail which has a characteristic maximum at $T_s$, and is therefore related to the second order structural phase transition. However, it does not show a tendency to go to zero at low temperatures. Fig.~\ref{differential} (c) shows the temperature evolution of the frequency of the most intensive part of the hump.

\section{Discussion}

At first, two issues of our observations should be addressed: (i) the strong hardening of the B$_{1g}$(Fe) phonon frequency which is more than two times stronger compares to the expected one from lattice shrinking and the Gr\"uneisen law and also very different from the modest hardening of the A$_{1g}$(Se) phonon mode; and (ii) the existence of a very narrow phonon linewidth on the background of the strong and continuous phonon hardening, meaning that the mechanism of hardening does not contribute to the phonon decay. Note, that in the FeTe we observed a very conventional hardening of the B$_{1g}$(Fe) mode with $\gamma_{B1g(Fe)} = 2.15$ which is close to the expected $\gamma \sim 2$.~\cite{gnezdilov-11} Hence, there is no specific mechanism of hardening in the FeTe. At the same time both Raman active modes in FeTe demonstrate a very unconventional line width, one order larger compared to FeSe.

The difference between these two compounds is the spin state of iron. In Fe$_{1+y}$Te for temperatures below magnetic ordering the magnetic moment is around 2 $\mu_B$/Fe depending on the excess of iron content $y$.~\cite{wen-11} This magnetic moment supposes the iron spin state to be intermediate $S=1$ which is highly orbital frustrated. Indeed, a few types of alternating orbital ordering patterns, which are consistent with the magnetically ordered structure, have been found to occur in iron pnictide superconductors.~\cite{kruger-09} In our paper~\cite{gnezdilov-11} we argued that in FeTe the orbital frustration mechanism resulting from the Fe intermediate spin state creates an additional channel for phonon decay. As a proof of this we observed an almost doubling of the B$_{1g}$(Fe) phonon linewidth at the structural/magnetic phase transition at which strong orbital fluctuations are naturally expected. In contrast, in FeSe no visible response of linewidth of both B$_{1g}$(Fe) and A$_{1g}$(Se) phonon modes at the structural phase transition is observed. Thus, one can safely conclude that the orbital channel of phonon decay is not dominant in FeSe and, particularly, it is not relevant to the mechanism of the structural phase transitions in FeSe. Note, that at the $P4/nmm \rightarrow Cmma$ second order ferroelastic phase transition, which is realized in FeSe, the optic phonons are not affected and some acoustic mode demonstrates a soft mode behaviour through lowering of its sound velocity for wave vectors much lower than the Brillouin zone boundary. In FeSe this effect has been studied.~\cite{luo-12, luo-12b} Such kind of acoustic phonon softening also does not affect the standard lattice anharmonicity channel of optic phonon decay. Therefore the phonons linewidths remain unchanged through passing $T_s$ (see Fig.~\ref{fitparameters} c, d). The only manifestation of the structural phase transition in FeSe in our Raman spectra is seen in the continuous increase in scattering intensity below $T_s$ (see Fig.~\ref{fitparameters} e, f). This effect can be connected with some rearrangement in polarizability caused by orthorhombic distortions and its increase under lowering temperature.

Concluding the experimental observations, the above comparison of FeTe and FeSe phonon is compatible with an intermediate spin state for FeTe in the whole investigated temperature regime. As this state is unaltered it does not contribute to phonon hardening, while it is a source of strong orbital fluctuations. On the other hand, the mechanism of strong phonon hardening for FeSe does not contain frustrating orbital degrees of freedom. At low temperature, however, the iron ions should be in the low spin state which has no spin orbital frustration similar as the high spin states. We naturally arrive to the conclusion that at high temperatures the iron ions in FeSe reside in the high spin state and the source of the strong B$_{1g}$(Fe) phonon hardening is \textit{a dynamical crossover} from the high to low spin state with temperature lowering. The corresponding strong hardening of the high frequency part of the phonon spectrum in FeTe upon decrease of iron magnetic moment has already previously been found in our earlier studies.~\cite{pashkevich-12}

To further illustrate this interplay we performed ab-inito lattice dynamic calculations of FeSe under varying magnetic moments at the iron sites within the framework of density functional theory (DFT) and a frozen phonon approach. We applied the all-electron full-potential linearized augmented plane wave method (Elk code)~\cite{elkcode} with the local spin density approximation (LSDA)~\cite{perdew-92} for the exchange correlation potential and with the revised generalized gradient approximation of Perdew-Burke-Ernzerhof (PBEsol).~\cite{perdew-08} The calculations were performed on a $9\times9\times6$ grid which corresponds to 60 points in the irreducible Brillouin zone. We used the experimental unit cell parameters for FeSe in the tetragonal phase at 300 K~\cite{hu-11} at ambient pressure with the unit cell lattice constants fixed. To reproduce the fluctuating magnetic moments in the paramagnetic phase we replaced them by static ones and suggested some kind of magnetic order. The spin-polarized calculations were performed with and without structure optimization for M$_{Fe}=0$ and with structure optimization for M$_{Fe} = 2 \mu_B$. In the latter case a checkerboard type of antiferromagnetic ordering with magnetic moments along the $c$-axis was supposed. The results of our calculations and a comparison with the experiment are given in Table I.

\begin{table*} \caption{\label{tab:table1}Results of phonon-mode calculations in FeSe and a comparison with data from Raman scattering experiments.}
\begin{ruledtabular} \begin{tabular}{ cccccc }
 Modes&\multicolumn{2}{c}{Experimental data}&\multicolumn{3}{c}{$P4/nmm$ at 300 K}\\ \hline
 $P4/nmm$&295 K&7 K&$z_{Se} = 0.2624$&$z_{Se} = 0.2493$&$z_{Se} = 0.2363$\\
 setting& & &not optimized&optimized&optimized\\
  & & &M$_{Fe}$ = 0&M$_{Fe}$ = 2 $\mu_B$&M$_{Fe}$ = 0 \\ \hline
  E$_g$& & &124.3&131.6&143.0\\
  A$_{1g}(Se)$&179.8&182.5&178.3&204.7&210.6\\
  B$_{1g}$(Fe)&193.9&206.5&218.8&206.6&230.8\\
  E$_u$& & &231.6&255.3&269.3\\
  E$_g$& & &260.2&294.7&309.6\\
  A$_{2u}$& & &276.6&301.8&293.4\\
\end{tabular} \end{ruledtabular} \end{table*}

As we expected the results for optimized structures do not coincide with the experiments giving highly overestimated A$_{1g}$(Se) and B$_{1g}$(Fe) phonon frequencies for both M = 0 and M = 2 $\mu_B$ iron magnetic moments. This is due to a too low value of the calculated Se ion height ($z_{Se}$) that correspondingly decreases the Fe-Se distance compared to the experimental ones. However, these results simulate the phonon frequency behavior under decrease of iron magnetic moments. As one can see, the frequency of the B$_{1g}$(Fe) phonon drastically increases when the size of magnetic moments changes from M = 2 $\mu_B$ to M = 0, while the A$_{1g}$(Se) phonon frequency undergoes modest change. Note that we do not discuss here the question about the importance and necessity of spin-polarized calculations in iron HTSC to obtain true phonon frequencies and electron-phonon coupling constants that are now generally accepted~\cite{boeri-10} but we focus on the phonon energy dependence on the size of the magnetic moment.

The spin polarized calculations without structure optimization with M = 0 give much better coincidence with the experiment for the A$_{1g}$(Se) mode but again overestimate the B$_{1g}$(Fe) mode. This overestimation is expected, as the experimental value of $z_{Se}$ at 300 K is higher than the calculated $z_{Se}$ not only for M = 0 but also for static magnetic moment M$_{Fe}=2 \mu_B$. The high value of $z_{Se}$ corresponds to a high magnitude of the magnetic moment (see for instance~\cite{pashkevich-12}). Thus, one can conclude that under dynamical spin state fluctuations the iron paramagnetic moment at 300 K should exceed 2 $\mu_B$. This estimation is in qualitative agreement with probing of the local magnetic moment in the FeTe$_{0.3}$Se$_{0.7}$ by Fe $K_\beta$ x-ray emission spectroscopy.~\cite{gretarsson-11} At room temperature they found 2 $\mu_B$ for a fluctuating local magnetic moment in the paramagnetic phase. Note, that in the Raman studies of the superconductive Fe$_{1+y}$Te$_{1-x}$Se$_x$ compounds the B$_{1g}$(Fe) mode shows almost the same features of frequency hardening and linewidth narrowing as in our case.~\cite{um-12} This implies the same mechanism of this phenomena for these compositions.

Summarizing our treatment of experimental and calculated data on the width and anomalous hardening of the B$_{1g}$(Fe) phonon we conclude that the paramagnetic ground state in FeSe can be presented as a superposition of low and high iron spin states with larger contribution of the high spin state at room temperature. Under temperature lowering the comparative contribution of those states undergoes a rearrangement with a continuous growth of the low spin state part over the high spin state part. This is seen as a decrease of the ground state paramagnetic moment. A specific spin-state lattice coupling is contributing to the B$_{1g}$(Fe) phonon hardening which does not create an additional channel for phonon decay and linewidth increase. Our indirect observation of the Fe paramagnetic moment decrease is in nice agreement with the almost linear lowering of the paramagnetic spin susceptibility with decreasing temperatures in FeSe which has been observed in the NMR experiments.~\cite{imai-09} The idea of representing the magnetic ground state in iron based HTSC as a coherent superposition of different spin states has recently been proposed and discussed.~\cite{chaloupka-12} Here we conclude that the paramagnetic ground state in FeSe combines both low and high spin states and does not include the intermediate spin state.

Furthermore, the result of our calculation for non optimized structure with M = 0 at 300 K can be used to roughly estimate the expected phonon frequencies at low temperatures. Indeed, in first approximation one can neglect contributions to the hardening from thermal lattice contraction as the main contribution comes from the change of the iron spin state. Then supposing M = 0, as it should be expected at lowest $T$, we still have a disagreement between theory (218.8 cm$^{-1}$) and experiment (206.5 cm$^{-1}$) in the B$_{1g}$(Fe) phonon frequency at 7 K, supposing that M $\neq 0$. This is evidence for residual spin fluctuations (nonzero contribution of high spin state) which survive even at 7 K -- the lowest temperature of our experiment. In other words, FeSe remains in a paramagnetic metallic state upon approaching the superconductive state. Interestingly, this estimation has resemblance with results of an inelastic neutron scattering experiment in FeTe$_{0.35}$Se$_{0.65}$ where a small fluctuating magnetic moment was found below $T_c$.~\cite{xu-11} This system again shows similar features in the Raman spectra as observed in FeSe.~\cite{um-12}

Our most intriguing observation in FeSe is the onset and development of a highly polarized low frequency background below 150 K which becomes gap-like under temperature lowering. This feature exists only in the $ab$-polarization channel of scattering (B$_{2g}$-channel in the tetragonal setting). Meanwhile, we see an almost temperature independent background in all other polarizations (see Fig.~\ref{polarization}). To be sure with the sign and value of the respective Raman tensors we checked this response in $RR$ and $RL$ polarization geometry (not shown here) and no antisymmetric contribution into Raman tensors has been found. Thus we conclude that this background does not come from pure magnetic excitations or magnetic fluctuations. As seen in the Fig.~\ref{differential} c the quasi-elastic part of the $ab$-Raman spectra starts to develop around 150 K and reaches a characteristic maximum of intensity at $T_s$ with a following decrease towards low temperatures. Note, that the order parameter $\eta_S \sim u_{ab}$ (type of lattice distortion) of the tetra-ortho phase transition has the same symmetry as the excitations seen in the $ab$-polarization channel of scattering. Due to this symmetry and characteristic increase at $T_s$ we ascribe the quasi-elastic part of scattering to the fluctuation of the $u_{ab}$ lattice distortions. Its existence in the Raman spectra in a wide temperature regime far above and far below of $T_s$ points to a very unusual short-ranged lattice instability which persists even in the tetragonal phase. The rhombic instability of the tetragonal phase is usually considered a fingerprint of the electronic nematic fluctuations in underdoped iron HTSC.~\cite{fernandes-10}

The gap-like feature shows a gradual increase of the frequency of its most intensive spectral weight starting from 150 K and up to 60 K. The frequency becomes saturated at lower temperatures at the energy $\sim 450$ cm$^{-1}$. A very similar $\sim 290$ cm$^{-1}$ energy gap with a similar gap onset below 130 -- 140 K has been observed recently in FeSe using time-resolved optical spectroscopy.~\cite{wen-12} The gap size there was restored from indirect measurements and modeling which might explain the difference from our direct observation of the gap value.

The gap onset accompanies a striking reduction in optical reflectivity and an increase of the Hall coefficient~\cite{wen-12} both signaling strong alterations in the electronic orbital and spin subsystems of FeSe below 150 K at the Fermi level. Furthermore, the hump excitations we observe have the symmetry of transitions between manifolds of electronic intermediate spin states and low and high spin states, that are allowed to be observed in $ab$ polarization. Thus, the gapped manifold of the intermediate spin states as well as accompanying orbital degrees of freedom do not participate in the formation of the paramagnetic ground state. This might explain the absence of orbital frustration in the FeSe paramagnetic ground state which we detect as a very narrow phonon linewidth. Interestingly, some alteration in the whole spin states manifold as well as in the appearance of the spin state gap have been supported also by observation of an anomalous change of the slope in the temperature dependence of the Se $z$ coordinate that happens around 150 K.~\cite{horigane-09}

Our findings also lead back to the question of the magnetic origin of the structural tetra-ortho phase transition in the ``11'' family. In spite of its occurrence in FeSe without the onset of magnetic order like in FeTe, it occurs in FeSe on the background of some alteration in the magnetic subsystems. Note, in previous studies the existence of a structural phase transition similar to FeTe in nonmagnetic FeSe was the reason to decline the hypothesis about its magnetic origin.~\cite{wen-11}

\section{Summary}

Our experimental results as well as lattice dynamical calculations reveal a temperature evolution of the paramagnetic state in FeSe from the high iron spin state ($S=2$) to the low iron spin state. We establish a high sensitivity of the B$_{1g}$(Fe) phonon mode, that is a characteristic mode of the FeSe$_4$ tetrahedron layers, to the Fe spin state. Furthermore, we demonstrate the relevance of spin-orbital frustration despite a considerable metallicity of these systems. Our scenario and the comparison of FeSe with FeTe underlines the complexity of Fe high temperature superconductors with respect to their lattice and electronic degrees of freedom.

\begin{acknowledgments}
We thank I. Vitebskiy and A. Keren for useful discussions. This work was supported by NASU-RFBR Grant Nos. 01-02-12 and 27-02-12, the Russian Foundation for Basic Research through Grants, 12-02-90405, the German Science Foundation (DFG, SPP 1458) as well as by the State program of Ukraine for implementation and applications of grid technologies, Grant No. 232. The calculations were performed using the grid-cluster at DonFTI NANU.
\end{acknowledgments}

\end{document}